\documentclass[12pt,titlepage]{article}
\usepackage{epsfig}
\usepackage{graphicx}
\usepackage{bbold}

\def\beq{\begin{equation}}
\def\eeq{\end{equation}}
\def\bea{\begin{eqnarray}}
\def\eea{\end{eqnarray}}
\def\dCSA{p}

\def\J{{\mathcal J}}
\def\Vsphere{{\varpi}}
\def\LargeJ{{L}}

\begin{document}

\title{The compressibility of rotating black holes in $D$-dimensions}

\author{{Brian P. Dolan}\\
{\small Department of Mathematical Physics, National University
of Ireland,}\\
{\small  Maynooth, Ireland}\\
{\small and}\\
{\small Dublin Institute for Advanced Studies,
10 Burlington Rd., Dublin, Ireland}\\
{\small e-mail: bdolan@thphys.nuim.ie}}

\maketitle
\begin{abstract}

Treating the cosmological constant as a pressure, in the context of
black hole thermodynamics, a thermodynamic volume for the black hole
can be defined as being
the thermodynamic variable conjugate to the pressure, in the sense of a
Legendre transform. The thermodynamic
volume is explicitly calculated, as the Legendre transform of the 
pressure in the enthalpy,
for a rotating asymptotically anti-de Sitter Myers-Perry
black hole in $D$ space-time dimensions.  The volume obtained is 
shown to agree with previous calculations using 
the Smarr relation. The compressibility is calculated and shown to be non-negative and bounded. 

Taking the limit of zero cosmological constant, the compressibility of
a rotating black hole in asymptotically flat space-times is determined
and the corresponding speed of sound computed.  The latter is bounded above
and has an elegant expression purely in terms of the angular momenta,
in the form of quartic and quadratic Casimirs of the rotation group, $SO(D-1)$. 

\bigskip

\ 

\noindent
\hbox{Report No. DIAS-STP-13-08}\hfill \hbox{PACS nos: 04.60.-m;  04.70.Dy}
\end{abstract}

\section{Introduction}

The thermodynamics of black holes has been an active area of research ever
since Bekenstein and Hawking's seminal papers on the entropy and temperature
associated with the event horizon of a black hole, \cite{Bek,Haw}.  
Recently the r\^ole of pressure and volume has come under scrutiny
in this context. It was pointed out in \cite{KRT} that the presence
of a cosmological constant, $\Lambda$,  
spoils the otherwise successful Smarr relation
\cite{Smarr} and a remedy was proposed: to raise $\Lambda$ 
to the status of a thermodynamic
variable, on a par with the temperature, while at the same time the  
black hole mass should be interpreted as the thermodynamic potential
associated with the enthalpy, rather than the heretofore
more usual interpretation of internal energy.
It is then very natural
to identify $\Lambda$ as being proportional to a pressure
and the thermodynamic variable conjugate
to the pressure can be interpreted as a volume for the black hole
\cite{Volume}. 
The idea of promoting $\Lambda$ to the status of a thermodynamic variable is not
new, \cite{HenneauxTeitelboim}-\cite{LPPV}, but it is only recently that
a volume has entered the picture in this context. 
For a rotating black hole in four dimensions
this thermodynamic volume does not have an
obvious relation to any geometric volume, though they agree if the black hole
is not rotating, \cite{VolumeJ}.
Nevertheless, with the volume included, there is a remarkable similarity between the black hole
equation of state and that of a Van der Waals gas, \cite{VolumeJ}-\cite{AKMS}.

We cannot vary $\Lambda$ in our Universe but we can nevertheless
gain considerable conceptual insight into the nature of black hole thermodynamics by performing a gedanken experiment performing such a variation.
Varying $\Lambda$ can have definite consequences: treating it as
a thermodynamic variable, interpreted as a pressure, introduces a $PdV$ term into the first law of black hole thermodynamics which affects the efficiency of
a Penrose process. In 4-dimensions, for example, the maximal efficiency
for energy extraction from a rotating electrically neutral black hole is 
increased from 29\% in asymptotically flat space-time to 52\% in asymptotically anti-de Sitter space-time, while for an electrically charged black 
hole the corresponding figures are 50\% and 75\%, \cite{VolumeJ}.
Of course we do not live in anti-de Sitter space, so this observation is
unlikely to have observable consequences, just as the Hawking temperature has no observable consequences for astrophysical black holes. Nevertheless it is interesting conceptually to think about the possible consequences of such a variation.
For example a negative cosmological constant can stabilise an otherwise 
thermodynamically unstable
black hole, provided $|\Lambda|$ is of sufficient magnitude, leading to the well-known Hawking-Page transition \cite{HawkingPage}.  Also in the AdS/CFT correspondence approach to condensed matter systems \cite{Gubser}, a negative cosmological constant 
could generate a positive pressure in the condensed matter system,  a pressure
which in general we would wish to be able to vary.

Asymptotically anti-de Sitter space-time, in different dimensions, 
is also of central importance in the gauge gravity/correspondence \cite{AdSCFT} and its applications in non-abelian gauge theories.
A consequence of this correspondence is a relation between the value of the cosmological
constant (or, equivalently, the square of the anti-de Sitter curvature length $L$), the
Planck length $l_P$  and
the number of colours in the $SU(N)$ gauge theory, 
$(L/l_P)^4\approx N$, so varying $\Lambda$ is varying $N$, and $N\rightarrow \infty$ in the classical limit. As long as $N$ is very large there is nothing wrong
with considering it to be a continuous variable (this is exactly how the
number of particles in a gas is treated in the thermodynamic limit).
Insights into the gauge/gravity correspondence might well be gained by 
considering the 
number of colours to be a thermodynamic variable. 

An important physical quantity in any thermodynamics system that behaves
like a gas is the compressibility: low compressibility implies a stiff equation of state and a stable system while a large compressibility implies a soft
equation of state and a system which is verging on instability.
Investigating the compressibility of a black hole solution therefore offers the possibility of learning something about stability without leaving the parametric framework of the solution.

The compressibility of rotating, asymptotically anti-de Sitter (AdS)
black holes in 4-dimensions was investigated in
\cite{Compressibility}, including asymptotically flat space-times as a
limiting case. In that work it was shown that the thermodynamic compressibility
of a rotating solar mass black hole in asymptotically flat space-time
is some 4 orders of magnitude less the that of a neutron star of similar mass,
sustained by degeneracy pressure.

In this paper the investigation of the
compressibility of rotating black holes is extended to 
dimensions greater than four, where it is known that there are instabilities
for ultra-spinning black holes.
To that end we first derive the compressibility of
 a rotating asymptotically AdS
Myers-Perry black hole in $D$ space-time dimensions.  
Our aim is to derive the compressibility and the speed of sound for asymptotically flat Myers-Perry black holes, but we must include a non-zero $\Lambda$ 
in order to obtain the volume and the compressibility before taking the
limit $\Lambda\rightarrow 0$.
In this limit
the expressions simplify considerably
and the compressibility and the speed of sound can
be expressed rather compactly in terms of the quadratic and quartic Casimirs of $SO(D-1)$
associated with the angular momenta of the black hole.

As was emphasised in \cite{VolumeJ}, 
it is crucial that one starts with rotating black holes: if there is no rotation
the entropy $S$ and the volume $V$ are both functions of the event horizon radius
$r_h$ only --- they are not independent and cannot be considered to be independent
thermodynamic variables --- they become independent only when the black hole rotates.
This is not a pathology it is merely due to the fact
 the Legendre transform is not well defined in the limit of zero rotation. 
The enthalpy, $H(S,P,J)$, is linear in $P$ when $J=0$ and hence the Legendre transform
is not invertible in this limit.  Everything is well defined as $J\rightarrow 0$,
provided one takes all derivatives and performs any desired Legendre transforms first
before setting $J=0$. Setting $J=0$ at the 
start misses crucial aspects of the thermodynamics as far as the thermodynamic
volume is concerned, but everything is consistent if the $J=0$ case is always
approached by setting $J\ne 0$ first and only letting $J\rightarrow 0$ after any
necessary manipulations involving Legendre transforms have been done. 

It is well known that Kerr black holes are thermodynamically unstable, indeed all of the rotating black hole space-times considered here are thermodynamically unstable in the asymptotically flat limit, \cite{DFMR}.
In $D=4$ the Schwarzschild black hole can be stabilised by placing it in a cavity with a heat bath 
at a finite distance from the horizon \cite{York} or by switching on a negative cosmological constant \cite{HawkingPage} of sufficient magnitude.  
A negative cosmological constant can also stabilise rotating black holes in $D>4$ though,
 while it may be possible to stabilise
rotating black holes by using a rotating cavity of finite size, it seems unlikely that one could ignore back-reaction and recover an asymptotically flat space-time if the radius of the cavity is sent to infinity.  Nevertheless one can gain significant insight into black hole thermodynamics by analysing thermodynamically unstable black holes, indeed this is how Hawking 
made the discovery that black holes could evaporate. 
In this work we shall only consider isolated asymptotically AdS black holes (which can be stable if the magnitude of $\Lambda$ is large enough) and their asymptotically flat limits.

An issue that should be addressed in discussing black hole thermodynamics is the r\^ole of extensive versus intensive variables.  In standard thermodynamics it is clear which variables scale with the volume of the system and which do not.  In black hole thermodynamics, with $c=1$, all relevant thermodynamic variables can be made to scale with dimensions of length to some power by judicial use of Newton's constant.  In $D$ space-time dimensions the key players for an electrically neutral black hole, which is the case for the analysis presented here, have the following length
dimensions :
\medskip

\setlength{\tabcolsep}{15pt}
\begin{tabular}{| l| c| c| }
\hline
 Mass, $M$ (enthalpy, $H$)   &  $D-3$ \\
  Entropy, $S$ (area)  & $D-2$ \\
  Angular momenta, $J^i$ & $D-2$ \\
 Volume,  $V$ & $D-1$ \\
 Temperature,  $T$ & $-1$ \\
 Angular velocity,  $\Omega_i$ & $-1$ \\
 Pressure, $P$ (Cosmological constant, $\Lambda$) & $-2$ \\
\hline
\end{tabular}
\medskip

\noindent Identifying the mass with the enthalpy $M=H(S,J,P)$ gives
\beq d H = T d S + \Omega_i\, dJ^i + V d P \eeq
and scaling produces the Smarr relation
\beq (D-3)M = (D-2)  S T + (D-2) \Omega_i J^i -2 PV.\eeq

It is thus very natural to interpret the variables whose scaling depends on the
space-time dimension $D$ as extensive and those whose dimension is independent 
of $D$ as intensive, and this will be the point of view adopted in this work.
Thus $H$, $S$, $J^i$ and $V$ are considered to be extensive, 
while $T$, $\Omega_i$ and $P$ are intensive.
The internal energy $U(S,J,V)= H-PV$ is thus a function of purely extensive variables,
\cite{VolumeJ}, while the enthalpy is not.

We restrict the analysis here to asymptotically AdS and asymptotically flat space-times.  The thermodynamics of black holes in $\Lambda > 0$ space-times
is a notoriously delicate issue.  First steps in understanding the r\^ole of
a thermodynamic volume of black holes in this case were
taken in \cite{MPdS} but unresolved issues remain, these 
are left for future work and are avoided here
by restricting to $\Lambda\le 0$.

In section \S\ref{AdSMPholes} we summarise the relevant features of 
asymptotically AdS Myers-Perry black holes, determine the thermodynamic
volume and describe the compressibility, the main result is the
compressibility in equation (\ref{eq:kappa}). In \S\ref{MPholes} the
$\Lambda\rightarrow 0$, asymptotically flat, limit is investigated;
the compressibility, given in (\ref{eq:kappaZero}) and speed of sound in
(\ref{eq:SpeedOfSound}), are derived and physical implications are discussed, particularly
in relation to ultra-spinning black holes in $D>4$.
The conclusions discuss some implications of the results and possible
future directions.  Finally some technical details are confined to two appendices.  

\section{AdS Myers-Perry black holes\label{AdSMPholes}}

Rotating black holes in $D$-dimensions must be treated slightly
differently for even and odd $D$
because the rotation group  $SO(D-1)$, 
acting on the event horizon which is assumed to have
the topology of a $(D-2)$-dimensional sphere, has different characterisations 
of angular momenta in the even and odd 
dimensional cases.
The Cartan sub-algebra has dimension $\frac{D-2}2$ for even $D$ and $\frac{D-1}2$ for odd $D$ so a general state of rotation is specified by 
$\frac{D-2}2$ independent angular momenta in even $D$ and $\frac{D-1}2$
in odd $D$.  Let $\dCSA=\left\lfloor \frac{D-1}2 \right\rfloor$, the integral part
of $\frac{D-1}2$, be the dimension of the Cartan sub-algebra of $SO(D-1)$,
then there are $\dCSA$ independent angular momenta $J_i$,
$i=1,\ldots,\dCSA$.  It is notationally convenient to define $\epsilon=\frac{1+(-1)^D}{2}$, so $\epsilon=1$
for even $D$ and $\epsilon=0$ for odd $D$, and then \beq \dCSA=\frac{D-1-\epsilon}2.\eeq

In this notation the unit $(D-2)$-dimensional sphere can be described in
terms of Cartesian co-ordinates $x_a$ in
${\bf R}^{D-1}$ by
\beq \sum_{a=1}^{D-1} x_a^2 =1,\eeq
and we can write this as
\beq \sum_{i=1}^\dCSA \rho_i^2 + \epsilon y^2=1,\eeq
where $x_{2i-1}+ix_{2i} = \rho_i e^{i\phi_i}$, $i=1,\ldots,p$, are complex co-ordinates
for both the even and odd cases while $y=x_{D-1}$ is only necessary for even $D$.
$\rho_i$, $\phi_i$ and $y$ are then (redundant) co-ordinates that can be used
to parameterise the sphere and, for the black hole, $J_i$ are angular
momenta in the $(x_{2i-1},x_{2i})$-plane.

The first rotating black solutions to Einstein's equations
in dimension greater than four were
were the asymptotically flat solutions of Myers and Perry \cite{MyersPerry}.
Rotating black holes in $5$-dimensions with a cosmological constant,
$\Lambda$, were constructed in \cite{HHT-R} and the generalisation to
the $D$-dimensional
metric was found in \cite{GPP}:
they are solutions of Einstein's equations with Ricci tensor\footnote{We use
units with Newton's constant and the speed of light set to unity,
$G=c^2=1$.}
\beq
R_{\mu\nu}=\frac{2\Lambda}{(D-2)}g_{\mu\nu}.
\eeq 
We shall focus on $\Lambda\le 0$ here, as the thermodynamics is then
better understood, and for notational convenience we define
\beq \lambda=-\frac{2\Lambda}{(D-1)(D-2)}\ge 0.\eeq
The line element in \cite{GPP} can then be expressed, in
Boyer-Linquist co-ordinates, as\footnote{The form given here differs
slightly from that in \cite{GPP} in that our ordinates, $t$ and $\phi_i$,
are related to those of \cite{GPP}, $\tau$ and $\varphi_i$, by 
$d\tau=dt$ and $d\phi_i = d\varphi_i -\lambda a_i dt$.}
\bea
d s^2&=&-W(1+\lambda r^2)dt^2 +\frac {2\mu }U
\left(W dt-\sum_{i=1}^\dCSA  \frac{a_i \rho_i^2d\phi_i}{1-\lambda a_i^2}\right)^2
\nonumber\\
&& +\left(\frac {U }{Z-2\mu}\right)d r^2 + \epsilon\, r^2 d y^2 
 +\sum_{i=1}^\dCSA\left(\frac{r^2+a_i^2}{1-\lambda a_i^2}\right)(d\rho_i^2+\rho_i^2d\phi_i^2 )\\
&&-\frac{\lambda}{W(1+\lambda r^2)}
\left(\sum_{i=1}^\dCSA \left(\frac{r^2+a_i^2}{1-\lambda a_i^2}\right)\rho_i d\rho_i +
\epsilon r^2 ydy\right)^2,
\label{eq:GPPds}
\nonumber\eea
where the functions $W$, $Z$ and $U$ are 
\bea W&=& \epsilon y^2+\sum_{i=1}^\dCSA \frac{\rho_i^2}{1-\lambda a_i^2}\nonumber\\
Z&=&\frac{(1+\lambda r^2)}{r^{2-\epsilon}}\prod_{i=1}^\dCSA(r^2+a_i^2)\\
U&=&\frac{Z}{1+\lambda r^2}\left(1-\sum_{i=1}^\dCSA\frac{a_i^2\rho_i^2}{r^2+a_i^2} \right).\nonumber\eea
The $a_i$ are rotation parameters in the $(x_{2i-1},x_{2i})$-plane,
restricted to $a_i^2<1/\lambda$, and $\mu$
is a mass parameter.

Many of the properties of the space-time with line element
(\ref{eq:GPPds}) were described in \cite{GPP}. 
There is an event horizon at $r_h$, the largest root of $Z-2\mu=0$, 
so
\beq 
\mu=\frac{(1+\lambda r_h^2)}{2r_h^{2-\epsilon}}\prod_{i=1}^\dCSA(r_h^2+a_i^2),
\label{eq:mudef}\eeq
with area
\beq 
{\cal A}_h=\frac{\Vsphere}{r_h^{1-\epsilon}}\prod_{i=1}^\dCSA\frac{r_h^2+a_i^2}{1-\lambda a_i^2}\,,
\label{eq:Adef}
\eeq
where $\Vsphere$ is 
is the volume of the round unit $(D-2)$-sphere,
\beq
\Vsphere= \frac{2\pi^{\frac{(D-1)}{2}}}{\Gamma\left(\frac{D-1}{2} \right)}\,.
\eeq

The Bekenstein-Hawking entropy is
\beq 
S=\frac{\Vsphere}{4 r_h^{1-\epsilon}}\prod_{i=1}^\dCSA\frac{r_h^2+a_i^2}{1-\lambda a_i^2}
\label{eq:Sdef}\eeq
and the Hawking temperature is, with $\hbar=1$,
\beq
T=\frac{r_h}{2\pi}(1+\lambda r_h^2)\sum_{i=1}^\dCSA\frac{1}{r_h^2 + a_i^2} + \frac{(2-\epsilon)(\epsilon\lambda r_h^2-1)}{4\pi r_h}.
\eeq

The  angular momenta and the ADM mass, $M$, of the black hole are related to the metric parameters via
\bea  J_i &=& 
\frac{\mu\,\Vsphere a_i}{4\pi(1-\lambda a_i^2)\prod_j (1-\lambda a_j^2)},   \label{eq:Jdef} \\
M&=&\frac{\mu\,\Vsphere}{8\pi\prod_j (1-\lambda a_j^2)} 
\left(D-2+2\lambda\sum_{i=1}^\dCSA\frac{a_i^2}{1-\lambda a_i^2}\right)
\label{eq:Mdef}\\
&=&\frac{(D-2)\mu\,\Vsphere}{8\pi\prod_j (1-\lambda a_j^2)}+\lambda\sum_{i=1}^\dCSA J_i a_i ,\nonumber
\eea 
while the angular velocities are
\beq \Omega_i=\frac{(1+\lambda r_h^2)a_i}{(r_h^2+a_i^2)}.\eeq

It was argued in \cite{KRT} that, in the presence of a cosmological constant,
the correct thermodynamic interpretation of the black hole mass is
that it is  the enthalpy of the system 
\beq M=H(S,P,J),\eeq
where $J$ stands for all the $J_i$ collectively, and the pressure is 
\beq P=-\frac{\Lambda}{8\pi}= \frac{(D-1)(D-2)\lambda}{16\pi}.\eeq
The thermodynamic volume, $V$, is defined as the variable 
thermodynamically conjugate to $P$ \cite{Volume,VolumeJ},
\beq V=\left.\frac{\partial M}{\partial P}\right|_{S,J}
= \frac{16\pi}{(D-1)(D-2)}\left.\frac{\partial M}{\partial \lambda}\right|_{S,J}.
\eeq
Details of the calculation of the thermodynamic volume by this technique
are given in appendix \ref{ap:Volume} and here we quote the 
result (\ref{eqapp:V})
\bea
V&=&\frac{r_h {\cal A}_h}{D-1}\left\{1+\frac{(1+\lambda r_h^2)}{(D-2) r_h^2}\sum_{i=1}^\dCSA \frac{a_i^2}{(1-\lambda a_i^2)} \right\}\\
&=&\frac{r_h {\cal A}_h} {D-1}+\frac{8\pi}{(D-2)(D-1)} \sum_{i=1}^\dCSA a_i J_i\nonumber.\label{eq:Volume}
\eea
With the substitution $\lambda\rightarrow g^2$ this agrees with the result
\cite{CGKP}
for the black hole volume,
derived from the assumption that the Smarr relation, 
\beq 
(D-3)M= (D-2)T S +(D-2) \sum_{i=1}^p\Omega_i J_i -2PV,
\eeq 
holds. For $D=4$ it reproduces the corresponding expression in \cite{VolumeJ}.
With the substitution $\lambda\rightarrow - g^2$, (\ref{eq:Volume}) 
agrees with the black hole thermodynamic volume quoted in \cite{MPdS} 
for $\Lambda > 0$, again determined by assuming the Smarr
relation holds, but avoiding the complication of the existence of the 
cosmological horizon that is present in this case. 

It is now possible to define the adiabatic compressibility 
of the black hole \cite{Compressibility} as
\beq \kappa=-\frac{1}{V}\left.\frac{\partial V}{\partial P}\right|_{S,J}.\eeq
With the explicit form of the thermodynamic volume in (\ref{eq:Volume})
the compressibility can be computed using the technique outlined in appendix
\ref{ap:kappa}:
it evaluates to
\beq
\kappa=\frac{16\pi(1+\lambda r_h^2)}{(D-1)(D-2)^2}
\frac{\Bigl\{
\sum_{i=1}^\dCSA  \frac{a_i^4}{1-\lambda^2 a_i^4}
-\frac{1}{(D-2)}\Bigl(\sum_{i=1}^\dCSA \frac{a_i^2}{1-\lambda a_i^2}\Bigr)
\Bigl(\sum_{i=1}^\dCSA \frac{a_i^2}{1+\lambda a_i^2}\Bigr)
\Bigr\}}
{\Bigl\{r_h^2+\frac{(1+\lambda r_h^2)}{(D-2)} \sum_{i=1}^\dCSA 
\frac{a_i^2}{1-\lambda a_i^2}\Bigr\}
\Bigl\{1-\frac{2\lambda}{(D-2)} \sum_{i=1}^\dCSA 
\frac{a_i^2}{1+\lambda a_i^2}\Bigr\}}
\ .\label{eq:kappa}
\eeq


It is shown in the appendix that $\kappa\ge 0$. It is also not
difficult to prove that it is bounded above for $\lambda>0$:
to see this first observe that the denominator never vanishes because
\beq
D-2 -2\lambda\sum_{i=1}^\dCSA \frac{a_i^2}{1+\lambda a_i^2} =  \epsilon -1 +2\sum_{i=1}^\dCSA \frac{1}{1+\lambda a_i^2} \ge \frac{D-3-\epsilon}{2},
\eeq
with equality when all the $a_i$ achieve the maximum value, $a_1^2=\cdots a_\dCSA^2=\frac{1}{\lambda}$.  The numerator can diverge though, if any or all of
the $a_i^2$ approach $\frac{1}{\lambda}$, but when this happens the
first factor in curly brackets in the denominator also diverges, 
the singularities cancel and $\kappa$ remains finite.  For example, if
$m$ of the $a^2_i$ approach $1/\lambda$, with $1 \le m \le p$, 
and the others are all zero, then
\beq \kappa \rightarrow \frac{8\pi}{(D-1)(D-2)\lambda} = \frac{1}{2P},
\eeq
reflecting the fact that $V\propto \frac{1}{\sqrt{P}}$ in this limit.

A thermodynamic speed of sound, $c_s$, 
can be defined by using the homogeneous density,
\beq \rho = 
\frac{M}{V} 
=\frac{(D-1)(D-2)(1+\lambda r_h^2)}{16\pi r_h^2} 
\frac{\left(
1+\frac{2\lambda}{D-2}\sum_{i=1}^\dCSA \frac{a_i^2}{1-\lambda a_i^2}
\right)}
{\left(
1+\frac{(1+\lambda r_h^2)}{(D-2) \,r_h^2}
\sum_{i=1}^\dCSA \frac{a_i^2}{1-\lambda a_i^2} 
\right) }.\eeq
The usual thermodynamic relation can then be used to obtain a speed
of sound,
\beq
\frac{1}{c_s^2}=\left.\frac{\partial \rho}{\partial P}\right|_{S,J}= 1+\rho\kappa \ge 1.\label{eq:cs}
\eeq
so $0\le c_s^2\le 1$.
Again, for example, taking
$m$ of the $a^2_i$ to approach $1/\lambda$ with the others all zero,
\beq
\rho\rightarrow \frac{(D-1)(D-2)\lambda}{8\pi},
\eeq
and
$\rho\kappa \rightarrow 1$, giving $c_s^2=\frac 1 2$.

\section{Isentropic processes in asymptotically flat Myers-Perry space-times\label{MPholes}}

When the cosmological constant vanishes many of the expressions in
the previous section simplify considerably.
In particular 
\bea M&=&\frac{(D-2)\Vsphere\mu}{8\pi}, \qquad
S=\frac{\Vsphere}{4 r^{1-\epsilon}} \prod_{i=1}^\dCSA(r_h^2 + a_i^2)
=\frac{4\pi}{D-2} M r_h
\nonumber \\
J_i&=&\frac{2 M a_i}{D-2},
\qquad  \Omega_i=\frac{a_i}{r_h^2+a_i^2}.
\label{eq:MJOmega}\eea

In this section we shall focus on isentropic processes,
for which it is convenient to define the dimensionless
angular momenta
\beq
\J_i=\frac{2\pi J_i}{S}=\frac{a_i}{r_h}
\eeq
in terms of which the mass is
\beq
M=\frac{(D-2)}{16 \pi}\Vsphere r_h^{D-3} \prod_{i=1}^\dCSA(1+\J_i^2),
\label{eq:Massj}
\eeq
and the entropy is
\beq
S=\frac{\Vsphere}{4} r_h^{D-2}\prod_{i=1}^\dCSA(1+\J_i^2).
\label{eq:Sj}
\eeq

The thermodynamic volume is
\beq
V=\frac{r_h{\cal A}_h}{D-1}\left(1+\frac{\sum_i \J_i^2}{D-2} \right)
=V_0\prod_{i=1}^\dCSA (1+\J_i^2)\left(1+\frac{\sum_i \J_i^2}{D-2} \right),
\eeq
where $V_0=\frac{\Vsphere r_h^{D-1}}{D-1}$ is the volume of
an ordinary $D-1$ dimensional sphere of radius $r_h$.

The $\lambda\rightarrow 0$ limit of (\ref{eq:kappa}) is finite:
\beq
\kappa=\frac{16\pi r_h^2}{(D-1)(D-2)^2}
\left\{\frac{(D-2)\sum_i \J_i^4 -(\sum_i \J_i^2)^2}{D-2+\sum_i \J_i^2}\right\},
\label{eq:kappaZero} \eeq
with
\beq
r_h^2=\left\{ \frac{4 S}{\Vsphere \prod_i(1+\J_i^2)}\right\}^{\frac{2}{D-2}}\,.\label{eq:rhj}
\eeq
Thus at fixed $S$ the compressibility is simply expressible entirely in terms of quadratic and quartic Casimirs of $SO(D-1)$. 

For $D\ge 4$, $\kappa$ is positive, and it is identically zero in $D=3$:
this latter result is in intuitive accord with the
fact that gravity has no dynamics in the bulk in 3-dimensions, all of the
interesting physics is in the boundary conditions.

To understand $\kappa$ fully, it is necessary to take account of the constraints
imposed by the condition that $T\ge 0$.  For $\lambda=0$
\beq
T=\frac{1}{2\pi r_h}\left( \sum_{i=1}^\dCSA \frac{1}{1+\J_i^2} - 
1+\frac{\epsilon}{2}\right)
\qquad \Rightarrow \qquad
\sum_{i=1}^\dCSA \frac{1}{1+\J_i^2} \ge 1-\frac{\epsilon}{2}\, .
\label{eq:extreme}
\eeq
 For $D>4$ it is possible for some of the $\J_i$ to tend to infinity,
but not all of them --- the well
known phenomenon of ultra-spinning black holes.\footnote{Note that, although
an ultra-spinning black hole has a large $J_i$, the corresponding angular velocity need not be large: indeed $\Omega_i\rightarrow 0$ as $J_i\rightarrow\infty$.
The inverse of the isentropic momentum of inertia tensor is 
\beq {\cal I}^{-1}_{ij}=
\left.\frac{\partial \Omega_i}{\partial J_j}\right|_S=
\frac{1}{M r_h^2}
\left\{ 
\frac{(D-2)}{2} \frac{(1-\J_i^2)}{(1+\J_i^2)^2} \delta_{ij} 
+\frac{\J_i \J_j}{(1+\J_i^2)(1+\J_j^2)} 
\right\}.
\eeq
For large $\J_i$, ${\cal I}^{-1}$ develops a negative eigenvalue
and a negative moment of inertia
implies that $\Omega_i$ {\it decreases} as $J_i$ increases.
Indeed if one of the $\J_i$ tends to infinity as 
$\J_i=\LargeJ\rightarrow \infty$,
at constant finite $S$, then $r_h\approx \LargeJ^{-\frac{2}{D-2}}$,
from (\ref{eq:Sj}), and the corresponding element of 
$ {\cal I}^{-1}\approx-\LargeJ^{\frac{4}{D-2}-2} \rightarrow 0$ 
as $\LargeJ\rightarrow \infty$, provided $D>4$.
Ultra-spinning black holes do not have large angular momenta because they 
have large angular velocity, they have large angular momenta because their moment of inertia diverges as $J_i\rightarrow\infty$.}

Equation (\ref{eq:extreme}) says that the locus of allowed temperatures is thus bounded by hyperbolae in $\J$-space.
The case for $D=6$ is plotted in figure 1 below.  This is very similar
to plots in \cite{LivingReviews}, except there the $J_i$ are normalised using the appropriate power of the mass, relevant for isenthalpic processes, 
while here the entropy is used, for isentropic processes.

\begin{figure}[!h]
\centering\includegraphics[width=300pt,height=300pt]{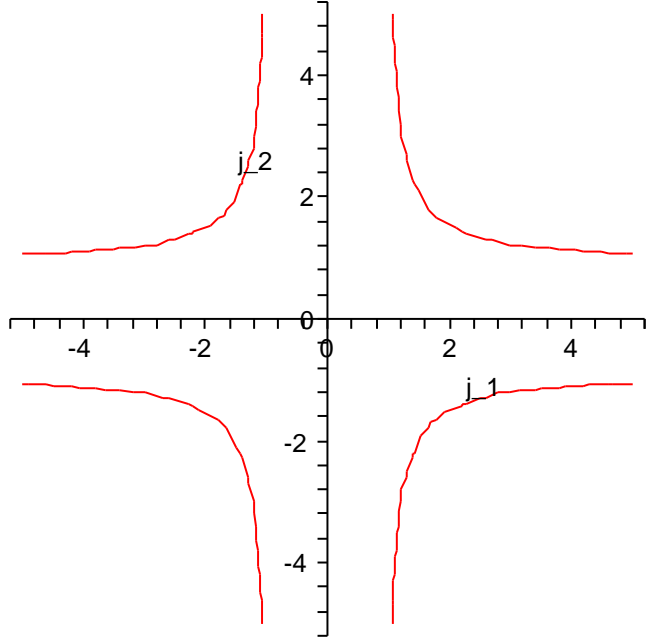}
\caption{The locus of extremal black holes, $T=0$, for $D=6$. $T>0$
requires the angular momenta to lie inside the region bounded by the hyperbolae.} 
\end{figure}

When all the $\J_i$ are small the compressibility is small
and the equation of state is very stiff, the black hole is
completely incompressible for $\J_i=0$. 
However  the compressibility can diverge if some $\J_i$ are kept small
while others are sent to infinity.  For example, if $\J_1=\cdots=\J_{p-m}=0$
and $\J_{p-m+1}=\cdots \J_p=\LargeJ$, then $T\ge 0$ for $\LargeJ\rightarrow\infty$ provided
$m\le\frac{D-3}{2}$. Also (\ref{eq:rhj}) implies that $r_h^2\propto \LargeJ^{-\frac{4m}{D-2}}$ so
\beq
\kappa\sim \LargeJ^{\frac{2(D-2m-2)}{D-2}},
\eeq
which diverges if $m < \frac{D-2}{2}$, so $\kappa$ diverges for $1\le m \le \frac{D-3}{2}$
with this configuration of angular momenta.  The divergence is fastest for
$m=1$.

When the compressibility becomes large the black hole
equation of state is very soft. 
For example the compressibility for $D=6$ is plotted in figure 2
and it grows indefinitely for large angular momenta along either the
$\J_1$ or the $\J_2$ axis, {\it i.e.}  $m=1$.
\begin{figure}[!h]
\centering\includegraphics[width=300pt,height=300pt]{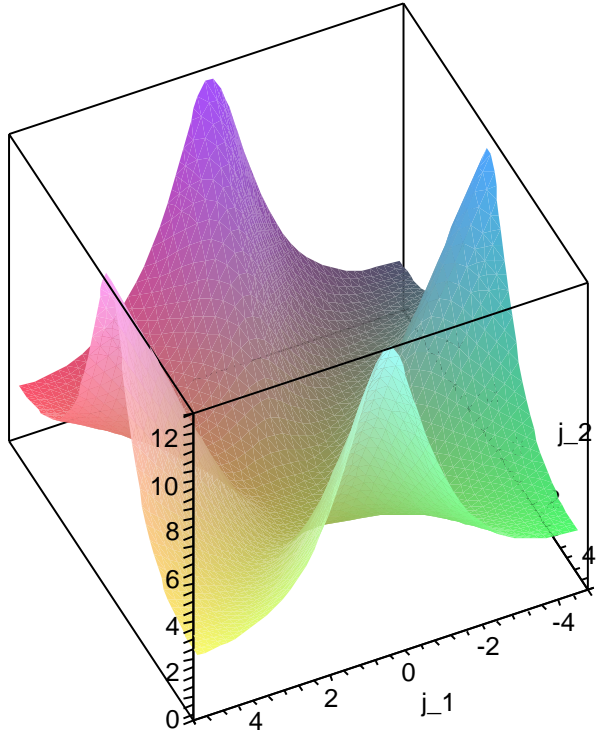}
\caption{The compressibility of a black hole in D=6 as a function of 
$\J_1$ and $\J_2$. }
\end{figure}

It was suggested in \cite{EM} that ultra-spinning black holes should
be dynamically unstable for large angular momentum, 
and subsequent numerical and analytical work supports this proposal 
\cite{EHNOR}-\cite{MPS}.  Large compressibility can be taken
as a sign of an instability setting in, although there is no indication
in equation (\ref{eq:kappaZero}) of a boundary in ${\cal J}$-space were a dynamical
instability might manifest itself, the expression for the compressibility
implies that the instability sets in more quickly when only one
angular momentum is taken to be large compatible with the pancake structure 
of \cite{EHNOR}.

Again a thermodynamic speed of sound can be defined using
\beq \rho= \frac{(D-1)(D-2)^2}{16\pi r_h^2}
\frac{1}{\bigl(D-2+\sum_{i=1}^\dCSA \J_i^2\bigr)}
\eeq
which gives, with  (\ref{eq:kappaZero}) in (\ref{eq:cs}),
\beq
c_s^2=\frac{1}{(D-2)} 
\frac{\bigl(D-2 + \sum_i \J_i^2\bigr)^2}{\bigl(D-2 + 2\sum_i \J_i^2 + \sum_i \J_i^4\bigr)}.\label{eq:SpeedOfSound} 
\eeq

It is not immediately clear 
how the thermodynamic speed of sound might
be related to a fluid dynamical speed of sound, but it is noteworthy that the
thermodynamic speed of sound is least when the compressibility is greatest,
as one would expect for a soft equation of state.
Indeed
$\frac{1}{D-2}\le c_s^2 \le 1$ with $c_s=1$ for $\J_i=0$ and 
$c_s^2\rightarrow \frac{1}{D-2} $ as any one $\J_i\rightarrow\infty$
with all others remaining finite.  It is possible that the thermodynamic
speed of sound is related to the velocity of the kind of waves and vibrations 
envisaged in \cite{EHNOR} associated with the instability of
an ultra-spinning black hole, at least for $D>4$.

\section{Conclusions \label{Conclusions}}

A cosmological constant spoils the Smarr relation for black hole
thermodynamics unless it is given the status of a thermodynamic
variable, most naturally interpreted as proportional to a pressure.
A consistent interpretation of the ADM mass of the black hole, in
terms of thermodynamic potentials, is that it is the enthalpy of the
black hole.  A thermodynamic volume can then be defined as being the
thermodynamic variable conjugate to the pressure, in terms of Legendre
transforms.  
 
The main results are the thermodynamic volume (\ref{eq:Volume}),
computed explicitly as the
Legendre transform variable 
conjugate of the pressure rather than by assuming the Smarr
relation, and the
compressibility (\ref{eq:kappa}) for Myers-Perry black holes
in asymptotically anti-de Sitter, $D$-dimensional, space-time.
The corresponding expressions for asymptotically flat space-times 
then follow easily from
the $\Lambda \rightarrow 0$ limit, and the corresponding quantities for
asymptotically flat Myers-Perry black holes are given in 
equations (\ref{eq:Volume}) and
(\ref{eq:kappaZero}) respectively.  In addition the speed of sound
can be expressed in terms of Casimirs of the rotation group, $SO(D-1)$,
and is given in (\ref{eq:SpeedOfSound}).

We emphasise again that it is crucial that the black hole is rotating.
It is clear from equations (\ref{eq:Adef}), (\ref{eq:Sdef}) and
(\ref{eq:Volume}) that,  when all $a_i\rightarrow 0$,
the entropy $S(r_h,\lambda,a_i)$ and the volume $V(r_h,\lambda,a_i)$ 
are both functions of the event horizon radius
$r_h$ only, we then have $V(r_h)$ and $S(r_h)$ and 
$V$ can be written uniquely as a function of the single variable $S$:
they cannot be considered to be independent
thermodynamic variables in this limit. The volume is an independent 
thermodynamic variable only when the black hole rotates, otherwise 
the Legendre transform is not well defined, as was first pointed out in
\cite{VolumeJ}.  This is reflected in the fact that the isentropic 
compressibility (\ref{eq:kappa}) vanishes as $\J_i\rightarrow 0$: fixing
$S$ fixes $V$ when the black hole is non-rotating, hence it is incompressible.

The discussion here has been restricted to electrically neutral rotating
black holes, leaving open the question of how electric charge might
affect compressibility.
 
It would be very interesting to develop these ideas in the context
of positive $\Lambda$, but we immediately hit the problem of having two
horizons to contend with, a black hole horizon and a cosmological horizon,
leading to two, in general different, temperatures and raising the question of
how to define thermodynamic potentials for such a system.  
A preliminary discussion of thermodynamic volumes in this context was given in \cite{MPdS}, but only by treating the two horizons 
as essentially independent and defining two independent volumes. The volume
associated with the black hole horizon in \cite{MPdS} 
was the same as (\ref{eq:Volume}), but
with $\lambda=-g^2$ negative, and it is perhaps significant in this
context that $\kappa$ in (\ref{eq:kappa}) remains positive under
this continuation to negative $\lambda$, provided $r_h^2<-\frac{1}{\lambda}$.
However a completely consistent integrated thermodynamic treatment of
asymptotically de Sitter space-times still eludes us.

\appendix

\section{Thermodynamic volume \label{ap:Volume}}

The thermodynamic volume is calculated by differentiating the mass 
(\ref{eq:Mdef}) with respect to $\lambda$, keeping the entropy and the angular momenta fixed.
To this end we note that
(\ref{eq:mudef}), (\ref{eq:Sdef}), (\ref{eq:Jdef}) and (\ref{eq:Mdef}) 
allow us to write
\beq 
J_i= \frac{S}{2\pi r_h} \frac{(1+\lambda r_h^2)}{(1-\lambda a_i^2)}\, a_i
\eeq
and demanding $dJ_i|_S=0$ then gives
\beq
d a_i =\frac{a_i}{(1+\lambda a_i^2)(1+\lambda r_h^2)}
\left\{(1-\lambda r_h^2)
\,(1-\lambda a_i^2)\, \frac{d r_h}{r_h} - (r_h^2 + a_i^2) \,d\lambda\label{eq:dai}
\right\}. 
\eeq
A second relation between $da_i$, $dr_h$ and $d\lambda$ follows from $dS|_{J_i}=0$ in (\ref{eq:Sdef}), allowing the elimination of $da_i$ to give
\beq
d r_h=\left(
\frac{\sum_i \frac{a_i^2}{1+\lambda a_i^2}}
{D-2+2\lambda\sum_i \frac{a_i^2}{1 + \lambda a_i^2}}  \right) 
r_h\, d\lambda,\label{eq:drh}
\eeq
and we have all the ingredients necessary to calculate 
$\left.\frac{\partial}{\partial\lambda}\right|_{S,J}$ acting on any function of $\lambda$, $r_h$ and $a_i$.

The thermodynamic volume is perhaps most easily calculated by combining 
(\ref{eq:mudef}), (\ref{eq:Sdef}), and the mass in
(\ref{eq:Mdef}), to write
\beq
M=\frac{S}{4\pi}\frac{(1+\lambda r_h^2)}{r_h}\left(D-2 + 2\lambda \sum_{i=1}^\dCSA \frac {a_i^2} {1-\lambda a_i^2} \right).
\eeq
Using this equation (\ref{eq:dai})  and (\ref{eq:drh}) yields the following formula for the volume
\bea
V=\frac{16\pi}{(D-1)(D-2)}\left.\frac{\partial M}{\partial \lambda}\right|_{S,J}
&=&\frac{4 r_h S}{D-1}\left\{1+\frac{(1+\lambda r_h^2)}{(D-2) r_h^2}\sum_{i=1}^\dCSA \frac{a_i^2}{(1-\lambda a_i^2)} \right\}\nonumber\\
&=&\frac{4 r_h S} {D-1}+\frac{8\pi}{(D-2)(D-1)} \sum_{i=1}^\dCSA a_i J_i.\label{eqapp:V}
\eea

\section{Compressibility \label{ap:kappa}}

The compressibility can be evaluated by pushing the analysis of appendix
\ref{ap:Volume} one step further and calculating
\beq \kappa = -\frac{16\pi}{(D-1)(D-2)}\frac{1}{V} \left.\frac{\partial V}{\partial \lambda}\right|_{S,J}
\eeq
A tedious, but straightforward calculation, gives
\beq
\kappa=\frac{16\pi(1+\lambda r_h^2)}{(D-1)(D-2)^2}
\frac{\Bigl\{
\sum_{i=1}^\dCSA  \frac{a_i^4}{1-\lambda^2 a_i^4}
-\frac{1}{(D-2)}\Bigl(\sum_{i=1}^\dCSA \frac{a_i^2}{1-\lambda a_i^2}\Bigr)
\Bigl(\sum_{i=1}^\dCSA \frac{a_i^2}{1+\lambda a_i^2}\Bigr)
\Bigr\}}
{\Bigl\{r_h^2+\frac{(1+\lambda r_h^2)}{(D-2)} \sum_{i=1}^\dCSA 
\frac{a_i^2}{1-\lambda a_i^2}\Bigr\}
\Bigl\{1-\frac{2\lambda}{(D-2)} \sum_{i=1}^\dCSA 
\frac{a_i^2}{1+\lambda a_i^2}\Bigr\}}
\ .\label{eqapp:kappa}
\eeq

We can show that $\kappa\ge 0$.
First note that
\beq
D-2 -2\lambda\sum_{i=1}^\dCSA \frac{a_i^2}{1+\lambda a_i^2} =  \epsilon -1 +2\sum_{i=1}^\dCSA \frac{1}{1+\lambda a_i^2} >0,
\eeq
hence both factors in curly brackets in the denominator of (\ref{eqapp:kappa}) are positive.
It remains to show that the curly bracket in the numerator is positive.
To this end define
\beq
X_i^\pm = \frac{a_i^2}{1\pm\lambda a_i^2}
\eeq
and express the curly bracket in the numerator in terms of the bi-linear form,
\beq
X^+.\,X^-:=\frac{1}{(D-2)}\sum_{i,j=1}^\dCSA X_i^+ K_{ij} X_j^-.
\eeq
$K_{ij}$ here are the components of the $\dCSA \times \dCSA$ matrix
\beq {\mathbf K} = (D-2){\mathbb  1} - {\mathbf I}\eeq
where ${\mathbf I}$ is the $\dCSA \times \dCSA$ all of whose entries are
1.
The eigenvectors of ${\mathbf K}$ are the same as the eigenvectors of ${\mathbf I}$ and the latter has one eigenvalue equal to $\dCSA$ and 
$p-1$ degenerate zero eigenvalues, hence ${\mathbf K}$ has one eigenvalue equal to
$D-2-\dCSA = \frac{D-3+\epsilon}{2}$ and $p-1$ eigenvalues equal to $D-2$.
All that concerns us here is that ${\mathbf K}$ is positive definite.
We can use the identity 
\beq
X^+.\, X^- = \frac{1}{2}\bigl( (X^+ + X^-).(X^+ + X^-)-X^+.\, X^+ - X^-.\,X^- \bigr),
\eeq
with $X_i^+ + X_i^-=\frac{2a_i^2}{1-\lambda^2 a_i^4}$,
to arrive at
\bea
X^+.\,X^-&=&\sum_{i,j=1}^\dCSA\frac{(1-\lambda^2 a_i^2 a_j^2)a_i^2 a_j^2}
{(1-\lambda^2 a_i^4)(1-\lambda^2 a_j^4)}K_{ij}
\nonumber \\
&\ge&
(1-\lambda^2 a_{\hbox{\tiny Max}}^2)
\sum_{i,j=1}^\dCSA\frac{a_i^2}{(1-\lambda^2 a_i^4)}
K_{ij}
\frac{a_j^2}{(1-\lambda^2 a_j^4)}, \nonumber
\eea
where $a_{\hbox{\tiny Max}}^2=\max(a_1^2,\ldots a_\dCSA^2)$.
Since all $a_i$ satisfy $a_i^2\le \frac {1}{\lambda}$ we have
$1-\lambda^2 a_{\hbox{\tiny Max}}^2\ge 0$ and hence $X^+.\, X^- \ge 0$.
The compressibility is thus bounded from below.

\end{document}